\documentclass{quantumview}
\usepackage[utf8]{inputenc}
\usepackage[english]{babel}
\usepackage[T1]{fontenc}
\usepackage{amsmath}
\usepackage{upgreek}
\usepackage{hyperref}
\usepackage{fixltx2e}
\usepackage[numbers,sort&compress]{natbib}
\usepackage{physics}

\begin{document}

\title{A projection-free approach toward mapping the structured polarization fields}
\author{Sandeep Singh*}
\affiliation{Photonic Sciences Lab., Physical Research Laboratory, Navrangpura, Ahmedabad 380009, Gujarat, India}
\affiliation{current affiliation: Center for Macroscopic Quantum States bigQ, Department of Physics, Technical University of Denmark, Kgs. Lyngby 2800, Denmark.}
\orcid{0000-0003-1533-8015}
\author{G.K. Samanta}
\affiliation{Photonic Sciences Lab., Physical Research Laboratory, Navrangpura, Ahmedabad 380009, Gujarat, India}
\maketitle

We present a projection-free method for mapping two-dimensional polarization distributions using Hong–Ou   
–Mandel (HOM) interference. Conventional polarization characterization techniques, such as Stokes polarimetry, rely on sequential intensity measurements under multiple polarization projections, making their accuracy and sensitivity susceptible to the extinction ratio, calibration errors, and stability of the polarization analysis optics. Our approach overcomes these limitations by exploiting the sensitivity of two-photon bunching to polarization indistinguishability at a balanced beam splitter. Using a HOM interferometer driven by a high-brightness spontaneous parametric down-conversion photon-pair source at 810 nm, we introduced a birefringent vortex waveplate with spatially varying polarization rotations in one interferometer arm and measured the resulting coincidence counts with a high signal-to-noise ratio, enabling projection-free characterization of the sample-induced polarization transformations. This configuration maps spatially dependent polarization variations directly onto coincidence counts, providing a projection-free reconstruction of the polarization distribution. We demonstrate high-fidelity ($\sim$95$\%$) reconstruction of the spatial polarization pattern with an angular resolution of approximately $\sim0.4^\circ$. The use an estimator that saturates the Cramer–Rao bound, computed from the Fisher information, can improve resolution further at the cost of longer acquisition. The proposed quantum-optical technique offers a simple, scalable, and high-precision framework for characterizing structured polarization fields in birefringent materials.
\section{Introduction}
Polarization is a fundamental degree of freedom of light, whose distribution encodes subtle information about both the field and the media it passes through, ranging from engineered optical components to complex biological tissues \cite{zhang2024}. Accurate characterization of these polarization patterns leads to critical advances in fields such as biomedical imaging\cite{ghosh2011}, material diagnostics\cite{tominaga2008}, quantum metrology\cite{han2023}, and secure quantum communication\cite{tan2024}. Traditionally, polarization characterization is performed using well-established Stokes polarimetry or its modified variants, which reconstruct the polarization state through sequential intensity projections with polarization-sensitive optics and time-averaged data acquisitions \cite{kumar2023, goldstein2017}. Although widely used, this technique is inherently indirect, relying on calibrated optical components in the analyzer arm that introduce losses, which are particularly non-negligible photons in photon-starved regimes\cite{zhang2024}, where extended integration is not feasible. Furthermore, the measurement sensitivity and accuracy are limited by the extinction ratio\cite{nabadda2024}, calibration errors, and long-term stability of the polarization analysis optics\cite{zhi2017}. Such conditions often arise with photosensitive media, such as biological tissues, where intensity-based methods also risk photo-damage \cite{waldchen2015}.

Unlike classical methods, quantum optics offers tools that probe polarization with enhanced sensitivity and minimal invasiveness by leveraging entanglement and quantum interference. Recent studies have explored polarization entanglement-based techniques that probe the sample with minimal invasiveness but still rely on rotating polarizers and coincidence measurements at multiple angles to extract polarization information \cite{zhang2024}. Alternatively, Hong–Ou–Mandel (HOM) interferometry is a technique that can be used to exploit the indistinguishability of the polarization of photon pairs to alter their joint probability detection \cite{singh2021, singh2023, Hong1987Mandel}. While HOM interference has been extensively explored in the temporal domain \cite{singh2021, singh2023} for high-precision parameter estimation \cite{singh2024fast}, its application in the polarization domain to probe polarization transformations remains comparatively underdeveloped \cite{polarization2020}. Recent work has demonstrated the feasibility of tracking linear polarization states by extracting Fisher information from HOM dip visibility and employing maximum-likelihood estimation \cite{polarization2020}. However, these approaches depend on extensive statistical post-processing and Cramér–Rao bound saturation, requiring a large number of measurement iterations to achieve adequate accuracy\cite{lyons2018, singh2023near}. The resulting long acquisition times restrict applicability to static or non-biological samples, as biological specimens degrade rapidly outside controlled environments \cite{mndlovu2024review}. In a recent proof-of-concept demonstration based on type-II spontaneous parametric down-conversion (SPDC) \cite{gonccalves2025}, the parametric gain of the crystal requires higher pump powers to improve the photon-pair generation rates at a cost of increased multi-photon events and accidental coincidences and thus sensitivity. Therefore, for improved estimation accuracy, it is essential to enhance the signal-to-noise ratio by improving the coincidence-to-accidental ratio (CAR), for example, by employing high-brightness photon-pair sources \cite{singh2022generation, dyer2009, Jabir:2017}.

Here, we demonstrate a fast, projection-free approach based on HOM interference for direct mapping of transverse linear polarization distributions of birefringent materials. Using a high-brightness spontaneous parametric down-conversion based pair photon source at 810 ± 2 nm, we directly map spatially varying polarization transformations induced by a birefringent vortex waveplate onto coincidence counts without the need for post-sample polarization projections. The high pair photon flux and large coincidence-to-accidental ratio (CAR) of the source enable rapid data acquisition and accurate reconstruction of transverse polarization distributions while simplifying the measurement architecture and minimizing systematic errors associated with conventional polarization-analysis techniques. These results establish HOM interferometry as a promising quantum-optical platform for high-precision characterization of structured polarization fields and other spatially varying birefringent media.

\section{Theory}
\subsection{Polarization depended rate of bunching}
\begin{figure}[ht]
    \centering
    \includegraphics[width=0.6\linewidth]{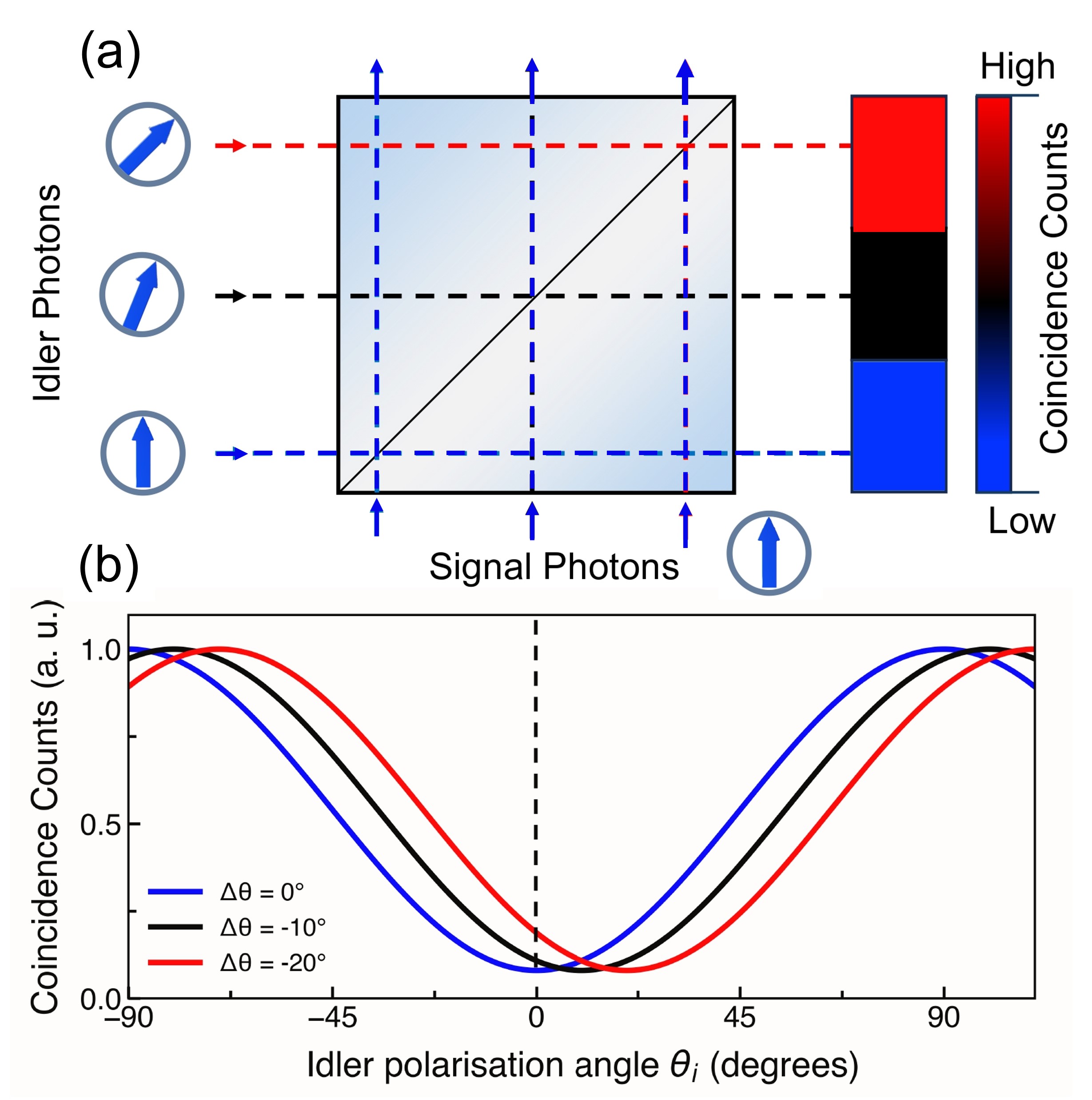}
    \caption{(a) Pictorial representation of change in coincidence counts with the relative change of polarization angle between signal and idler photons. (b) Variation of coincidence counts as a function of polarization rotation of the idler for different polarization angles of the signal from the vertical axis. The optical delay between the photons is zero. The vertical dotted lines represent vertical polarization of signal photons.}
    \label{Fig_1}
\end{figure}

The current concept, illustrated schematically in Fig. \ref{Fig_1}(a), is based on polarization-dependent Hong–Ou–Mandel (HOM) interference, where the photon-bunching probability at a balanced beam splitter depends critically on the polarization indistinguishability of the interfering photons. For a vertically polarized signal photon, interaction of the idler photon with a birefringent vortex waveplate introduces a position-dependent polarization rotation in the transverse plane. Consequently, different regions of the vortex waveplate produce different degrees of polarization distinguishability between the signal and idler photons, leading to corresponding variations in the HOM coincidence counts. Here, we consider three idler-photon polarization states corresponding to progressively increasing polarization rotations relative to the vertical polarization, represented by the blue, black, and red dotted lines. 

The coincidence counts exhibit the characteristic polarization-dependent HOM response, with minimum coincidence for parallel polarizations and maximum coincidence for orthogonal polarizations. As shown in Fig. \ref{Fig_1}(b), increasing the initial polarization rotation of the signal photon from $0^\circ$ (blue) to $-10^\circ$ (black) and $-20^\circ$ (red) results in minimum coincidence counts for the polarization rotation of the idler photon by the same amount, while preserving the shape and width of the HOM curve. This behavior effectively resets the polarization offset introduced by the sample. As a result, the change in coincidence counts along any vertical line in Fig. \ref{Fig_1}(b) directly reflects the relative polarization rotation between the signal and idler photons, providing a basis for mapping the transverse polarization distribution of the birefringent sample.

\subsection{Coincidence probability}
\label{sec_2.2}

The spontaneous parametric down-conversion (SPDC) process in a \(\chi^{(2)}\) nonlinear crystal generates frequency-entangled photon pairs under energy conservation \(\omega_p = \omega_s + \omega_i\) and phase matching conditions \cite{singh2023tolerance}. In the low-gain regime, for a narrowband pump, the output two-photon state can be written as follows:
\begin{equation}
|\Psi\rangle = \iint d\omega_s d\omega_i\,f(\omega_s,\omega_i)
\iint d\lambda_s d\lambda_i\,\,g(\lambda_s,\lambda_i)
\,\hat a_{1,V}^{\dagger}(\omega_s,\lambda_s)\,
\hat a_{2,V}^{\dagger}(\omega_i,\lambda_i)\,|0\rangle
\label{eq_1}
\end{equation}
where \(f(\omega_s, \omega_i)\) encodes spectral correlations and is commonly referred to as joint spectral amplitude (JSA), which includes pump and phase-matching properties. The terms $g(\lambda_s,\lambda_i)$ capture the correlation in residual degrees of freedom (DOFs), such as the spatial mode of the two-photon state. In the present experiment, the vertically polarized signal photon is sent through a temporal-delay stage, while the corresponding vertically polarized idler photon propagates through a birefringent object, such as a zero-order vortex half-wave plate (vortex waveplate), which modifies its polarization state. The polarization rotation angle of the idler photon with respect to the vertical axis is denoted as $\theta_{i}$, where positive values correspond to clockwise rotations. This angle is determined by the orientation of the fast axis of the vortex waveplate in the transverse position through which the idler photon propagates. The polarization of the idler photon after the sample can be written as,

\begin{equation}
|P\rangle_i = \cos\theta_{i}\,|V\rangle_i + \sin\theta_{i}\,|H\rangle_i
\label{eq_2}
\end{equation}
Using Eq. \ref{eq_1}, we can write the resulting biphoton state as:
\begin{equation}
|\Psi\rangle = \iint d\omega_s d\omega_i\,f e^{-i\omega_s t}
\iint d\lambda_s d\lambda_i\,g(\lambda_s,\lambda_i)
\left[\cos\theta_{i}\,\hat a_{1,V}^{\dagger}\hat a_{2,V}^{\dagger}
+ \sin\theta_{i}\,\hat a_{1,V}^{\dagger}\hat a_{2,H}^{\dagger}\right]|0\rangle
\label{eq_3}
\end{equation}
The 50:50 beam splitter used for interference transforms the input modes (\(1,2\)) to output modes (\(3,4\)) as per the following unitary transformation:
\begin{equation}
\hat{a}_{1,V}^\dagger(\omega) \rightarrow \frac{\hat{a}_{3,V}^\dagger(\omega) + \hat{a}_{4,V}^\dagger(\omega)}{\sqrt{2}}  ,\
\quad
\hat{a}_{2,P}^\dagger(\omega) \rightarrow \frac{\hat{a}_{3,P}^\dagger(\omega) - \hat{a}_{4,P}^\dagger(\omega)}{\sqrt{2}}
\label{eq_4}
\end{equation}
Substituting Eq. \ref{eq_4} into Eq. \ref{eq_3} and retaining only the post-selected terms corresponding to coincidence events, i.e., one photon emerging from each output port of the beam splitter, we obtain

\begin{multline}
|\Psi_{\text{coin}}\rangle = \frac{1}{2} \iint d\omega_s d\omega_i\,\, e^{-i\omega_s t} f(\omega_s, \omega_i)
\iint d\lambda_s d\lambda_i\,\, g(\lambda_s,\lambda_i) \\
\times \left\{\cos\theta_{i}\left[\hat a_{3,V}^{\dagger}\hat a_{4,V}^{\dagger} - \hat a_{4,V}^{\dagger}\hat a_{3,V}^{\dagger}\right]
+ \sin\theta_{i}\left[\hat a_{3,V}^{\dagger}\hat a_{4,H}^{\dagger} - \hat a_{4,V}^{\dagger}\hat a_{3,H}^{\dagger}\right]\right\}|0\rangle.
\label{eq_6}
\end{multline}
Assuming narrow-bandpass Gaussian filters centered at frequency $\omega_{\circ}$ at the output ports of the beam splitter, the coincidence probability, governing coincidence detection rates, can be calculated as follows.  \(P_c = \langle\Psi_{\text{coin}}|\Psi_{\text{coin}}\rangle\):

\begin{equation}
\begin{aligned}
P_c &= \frac{1}{4} \iiiint d\lambda_s \, d\lambda_i \, d\lambda_s' \, d\lambda_i' \,\, g^*(\lambda_s,\lambda_i) g(\lambda_s',\lambda_i') \,\times \\
&\iiiint d\omega_s \, d\omega_i \, d\omega_s' \, d\omega_i' \,\, f^*(\omega_s,\omega_i) f(\omega_s',\omega_i') \\
&\quad \times \,\, e^{i(\omega_i' - \omega_i)t} \,\Big[ (\cos^2\theta_{i} + \sin^2\theta_{i})\delta_{\text{d}} - \cos^2\theta_{i} \, e^{-2\sigma^2 t^2} \delta_{\text{e}} \Big]
\end{aligned}
\label{eq_7}
\end{equation}
where,
\begin{equation}
\delta_{\text{d}} \equiv \delta(\omega_s - \omega_s')\delta(\omega_i - \omega_i') \quad\text{\&}\quad 
\delta_{\text{e}} \equiv \delta(\omega_s - \omega_i)\delta(\omega_s' - \omega_i')
\end{equation}

For degenerate photons (\(\omega_s = \omega_i = \omega_{\circ}/2\)) with Gaussian spectra, the coincidence probability is expressed as
\begin{equation}
P_c(t, \theta_{i}) = \frac{1}{2} \left( 1 - \alpha\cos^2\theta_{i} \cdot e^{-2\sigma^2 t^2} \right)
\label{eq_8}
\end{equation}
here, $\alpha = \iint d\lambda_s d\lambda_i g^*(\lambda_s,\lambda_i)g(\lambda_i,\lambda_s)$, is defined as interference visibility and quantifies the mode-matching quality of the two-photon state. It is evident from Eq. \ref{eq_8} that the coincidence probability, $P_{c}$, and thus the interferometer's sensitivity, depend upon both the temporal $t$ and polarization rotation (distinguishability) $\theta_{i}$. The $\cos^2\theta_{i}$ term arises from polarization interference, whereas \(e^{-2\sigma^2t^2}\) reflects temporal mode overlap. Expressing \(t\) as an optical path difference \(x = ct\), the Eq. \ref{eq_8} can be re-written as;
\begin{equation}
P_c(x, \theta_{i}) = \frac{1}{2} \left( 1 - \alpha\cos^2\theta_{i} \cdot e^{-x^2/\sigma_{x}^2} \right), \quad \sigma_x = \frac{c}{\sqrt{2}\sigma}
\label{eq_9}
\end{equation}
where \(x\) signifies the path length (temporal) mismatch and \(\sigma_x\) is the full width at half maximum (FWHM) width of the interference dip. In an ideal scenario of zero optical delay \((x=0)\), the exponential term becomes unity and \(P_c(0,\theta_{i})=\tfrac12(1-\alpha\cos^{2}\theta_{i})\); hence, the interferometer acts as a pure polarization sensor. However, there will be some undesired fluctuations in $P_c(x, \theta_{i})$ due to small residual optical-path variations between the signal and idler arms across the raster scan, majorly arising from sample thickness inhomogeneity and mechanical drift of the translation stage mirror with amplitude, say $|{\delta x}|$, which perturbs this polarization-sensitive response, quantified as below:
\begin{equation}
\Delta P(x) \;=\; P_c(\delta x,0)-P_c(0,0)
           \;=\;\frac{\alpha}{2} (1-e^{-(\delta x/\sigma_x)^{2}})
\label{eq_10}
\end{equation}
We want to use the interferometer to measure only the changes in the polarization of the biphoton states; therefore, we tend to tune the interferometer to suppress its sensitivity to changes in $|{\delta x}|$. This can be achieved by making $\Delta P(x) \ \rightarrow 0$ as $\theta_{i} \rightarrow 0$, thus choosing the $\sigma_{x}$ broad enough such that $ \sigma_{x} \gg |\delta x| $.
\subsection{Optical-delay-induced polarization error}
\label{sec_2.3}
The experiment is designed to isolate polarization sensitivity, and is therefore analyzed under the simplified assumption of $x = 0$, corresponding to zero optical delay. However, in practice, translating the vortex waveplate across the transverse plane can also introduce unintended variations in the optical path length, $\delta x$, due to thickness inhomogeneity, thereby affecting the photon bunching, $P_{c}$. Because the $P_{c}$ has a coupled dependence on both optical delay and polarization distinguishability, such thickness-induced variations in the coincidence rate manifest as an apparent polarization rotation, which can be denoted by $\delta\theta_{i}$. This spurious rotation induces an error in the measured coincidence probability $\Delta P(x)$, which in turn leads to a systematic polarization measurement error, and can be quantified by analyzing the situation as $P_{c}(\delta x,0) = P_{c}(0,\delta\theta_{i})$
as follows:
\begin{equation}
\delta\theta_{i} \;=\; 2*\arccos(\exp({-\delta x}^{2}/2\sigma_{x}^2))
\label{eq_13}
\end{equation}
\section{Experimental Scheme}
\begin{figure}[h]
\centering
\includegraphics[width=0.8\linewidth]{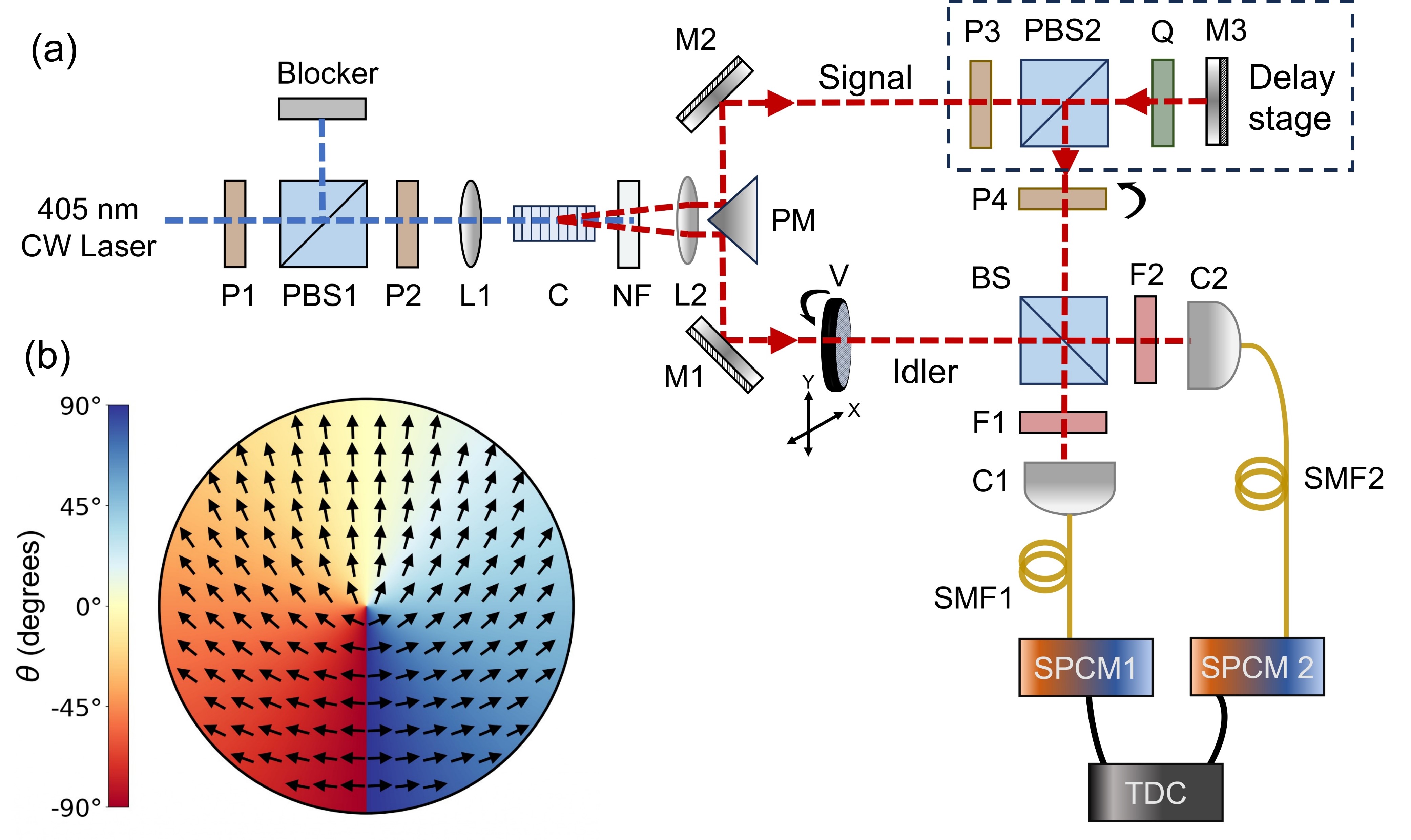}
\caption{(a) Optical layout of the experimental setup.  P1-3: half-wave plates; PBS: polarizing beam splitter cube; L1-2: plano-convex lenses, PPKTP: periodically poled KTP crystal; NF: Notch filter; PM: prism mirror, M1-i 3: dielectric mirrors, V: vector-vortex plate, Q: quarter wave plate, BS: 50:50 beam splitter cube, F1-2: interference filter; C1-2: fiber couplers, SMF1-2: single-mode fiber, SPCM1-2: single-photon counting modules; TDC: time to digital converter, (b) Spatial distribution of fast axis orientation of a zero-order vortex half-wave plate of m=1 (V) over its circular aperture.}
\label{Fig_2}
\end{figure}

The schematic of the experiment is shown in Fig.  \ref{Fig_2}. A continuous-wave (CW) diode laser centered at $\lambda_{p}$ = 405 nm, with a maximum output power of 100 mW and linewidth $<$ 8 MHz, is used as the pump laser. A combination of a half-wave plate (HWP, P1) and a polarizing beam splitter cube (PBS) is used to precisely control the pump power incident on the nonlinear crystal. The second HWP (P2) is used to adjust the polarization of the pump beam in the vertical direction, as required by the quasi-phase-matching criteria. A plano-convex lens (L1) with a focal length of $f_1$ = 100 mm focused the pump beam to a $\sim$40 $\upmu$m spot at the center of a periodically poled potassium titanyl phosphate (PPKTP) crystal (C), which served as the nonlinear medium. The crystal, with a length of 10 mm and an aperture of 1$\times$2 mm$^2$, is housed in a temperature oven with a temperature stability of $\pm$0.02$^{\circ}$C, to maintain quasi-phase-matching conditions.
The vertically polarized pump photons at 405 nm, upon interaction with the PPKTP crystal in a type-0 ($e \rightarrow e + e$) non-collinear phase-matching configuration, result in two degenerate down-converted daughter photons, conventionally termed the signal and idler. Owing to the position-momentum correlations, the downconverted photons exhibited a transverse profile in the form of an annular ring (commonly known as the SPDC ring)\cite{Jabir:2017}, with correlated photons located at diametrically opposite points on the ring. The downconverted photons are extracted from the residual pump beam using a notch filter (NF) centered at $\lambda_{p}$. A prism-shaped mirror (PM) bisected the SPDC ring such that, for each correlated pair, the signal and idler photons were directed into opposite arms of the interferometer, ensuring that they impinged on the input ports of the 50:50 beam splitter (BS) forming the HOM interferometer. A zero-order vortex half-wave plate (vortex waveplate, V) (Thorlabs) with a clear aperture diameter of 21.5 mm, a uniform half-wave retardance, and an azimuthally varying fast-axis orientation, as illustrated in Fig. \ref{Fig_2}(b), is placed in the reference arm of the interferometer. It is housed in a mechanical stage comprising a motorized rotating stage and a custom-built 2D motorized linear stage, enabling rotational control of its fast axis and raster scanning across its clear aperture. To vary the temporal overlap between the signal and idler photons, the signal photons passed through an optical delay stage comprising a HWP (P3), PBS, and a quarter-wave plate, QWP (Q). The P4, an HWP mounted on a rotational stage, is used to characterize the interferometer in the polarization degree of freedom. The downconverted photons emerging from the two output ports of BS were spectrally filtered by narrow-band (bandwidth $\sim$ 3.2 nm) interference filters (F1, F2) centered at 810 nm, and subsequently coupled into single-mode fibers using optical fiber couplers (C1, C2). The SMFs were connected to single-photon counting modules (SPCM, AQRH-14-FC, Excelitas), which were interfaced directly with a time-to-digital converter (TDC, ID800). Throughout this work, a temporal coincidence window of approximately 1.62 ns is used unless otherwise specified.

\section{Results and Discussions}
\subsection{Classical polarization projection}
\begin{figure}[h!]
    \centering
    \includegraphics[width=0.9\linewidth]{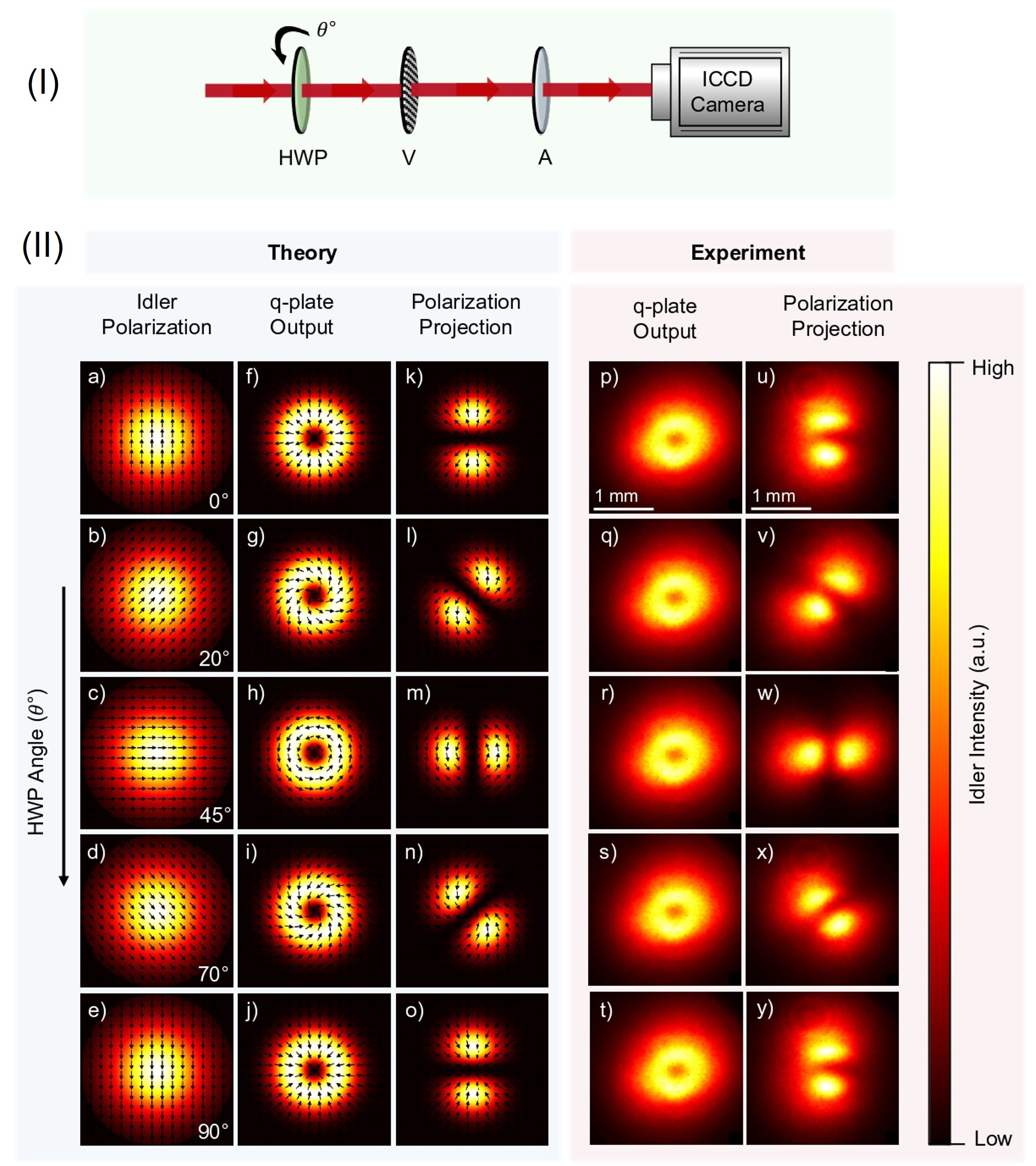}
    \caption{(I) working principle used for classical polarization projection. HWP: Half-wave plate, V: vortex waveplate, A: Analyzer. (II)) Evolution of a linearly polarized Gaussian beam through a first-order $q$-plate and its polarization projections. Left panel (Theory): (a–e) Input idler polarization states for half-wave plate angles $\theta = 0^\circ, 20^\circ, 45^\circ, 70^\circ, 90^\circ$. (f–j) Corresponding $q$-plate output showing transverse intensity and polarization distribution. (k–o) Simulated intensity after vertical polarization projection. Black arrows denote local polarization orientations. Right panel (Experiment): (p–t) Measured intensity distribution of the idler photons after propagation through the $q$-plate (V), and corresponding (u–y) polarization projection of the photons after the analyzer, A. }
    \label{Fig_3}
\end{figure}
\noindent
To get a better perspective on the current work, we first illustrate the concept of conventional polarization projection using the idler photons taken out from the experimental setup shown in Fig. \ref{Fig_2}. In the absence of heralding, the photons in the idler arm exhibit thermal-light statistics and may be treated as a classical electromagnetic field\cite{caspani2017,mandel1996}. The experimental arrangement, shown in Fig. \ref{Fig_3}(I), consists of a half-wave plate (HWP) mounted on a rotation stage for variable fast axis rotation angle ($\theta^\circ$), followed by a birefringent vortex waveplate and a polarization analyzer (A) fixed at a chosen projection angle. The analyzer is essential and must be placed in the idler beam path to perform projection measurements, as it is a standard practice in polarization projection studies \cite{he2021polarisation, kumar2023}. The transmitted photons are subsequently recorded using an intensified CCD (ICCD) camera to obtain the spatial distribution of the photons. 
To simplify the understanding, the idler beam is modeled as a Gaussian mode with uniform linear polarization, controlled using the HWP. The corresponding electric field amplitude in cylindrical coordinates is given by;
\begin{equation}
\mathbf{E}(r,\theta) = E_0 \exp\left(-\frac{r^2}{w_{o}^2}\right) (\hat{\vb{e}}_x \sin\theta + \hat{\vb{e}}_y \cos\theta)
\label{eq:gaussian}
\end{equation}
Here, $w_0$ denotes the $1/e$ beam waist, and $(r, \theta)$ are cylindrical coordinates. We adopt the sign convention that the polarization direction, denoted by, $\theta$ is defined with respect to the $y$-axis and measured positively in the clockwise direction. $\hat{\vb{e}}_x$ and $\hat{\vb{e}}_y$ are the unit vectors along $x$ and $y$ directions, respectively. The first column, Fig. \ref{Fig_3}(I)(a–e), shows the calculated spatial intensity distributions and corresponding polarization directions (black arrows) of the input beam after the HWP for different rotation angles, $\theta^o$. In all cases, the polarization remains uniform across the transverse plane, yielding a spatially homogeneous polarization state. Subsequently, the beam traverses the first-order $q$-plate with the fast-axis orientation, $\theta_{\text{f}} = \frac{m}{2}\phi + \delta$, where $\phi = \tan^{-1}(x',y')$ is the azimuthal angle in the transverse plane, and $\delta$ denotes the residual polarization offset\cite{cardano2012}. The polarization of the input beam after passing through the vortex waveplate is rotated by an angle, $\theta_{\text{i}}$ = 2($\theta_{\text{f}}$-$\theta$), where $\theta_{\text{f}}$ is the orientation of the vortex waveplate fast axis. Consequently, the spatial electric field distribution of the output photons can be expressed as\cite{karimi2007}
\begin{equation}
\mathbf{E}_{\text{q}}(r,\phi) = \sqrt{\frac{2^{|m|+1}}{\pi w_{o}^{2} |m|!}}\left(\frac{r}{w_{o}}\right)^{|m|} \exp\left(-\frac{r^2}{w_{o}^2}\right) (\hat{\vb{e}}_x \sin\theta_{i} + \hat{\vb{e}}_y \cos\theta_{i})
\label{equ_13}
\end{equation}
This transformation yields a characteristic doughnut-shaped intensity profile\cite{kumar2022} with spatially varying polarization, as seen in the second column (f-j) of Figs. \ref{Fig_3}(II). Although the intensity patterns of the output photons are invariant to the input polarization state, the local polarization orientation, as shown by the black arrows, gets modified with $\theta$. 
Subsequently, the polarization projection of the photons using the analyzer with the transmission axis set along $\hat{\vb{e}}_y$ are shown by the third column, (k-o) of Fig. \ref{Fig_3}(II). The projected intensity can be mathematically represented as, 
\begin{equation}
I_{\text{v}}(x',y') = \left| \vb{E}_{\text{q}} \cdot \hat{\vb{e}}_y \right|^2 = \left|\mathbf{E}_{\text{q}}(r,\phi)\right|^2 \cos^2(m\phi + 2\delta)
\label{equ_14}
\end{equation}
The measured spatial intensity distributions at the output of the vortex waveplate of order, m = 1 and the corresponding projected intensity distributions after the analyzer for different input polarization angles are shown in the fourth column, Fig. \ref{Fig_3}(II)(p–t), and the fifth column, Fig. \ref{Fig_3}(II)(u–y), respectively. The weak background observed in the experimental images can be attributed to the residual birefringence of the vortex waveplate, which was designed for operation at 1064 nm but is used here at 810 nm\cite{kadiri2019}. Despite this background contribution, the experimental results are in excellent agreement with the theoretical predictions, confirming the polarization transformations induced by the vortex waveplate and their subsequent projection by the analyzer. 

\subsection{HOM interference}

\begin{figure}[h!]
    \centering    
    \includegraphics[width=0.9\linewidth]{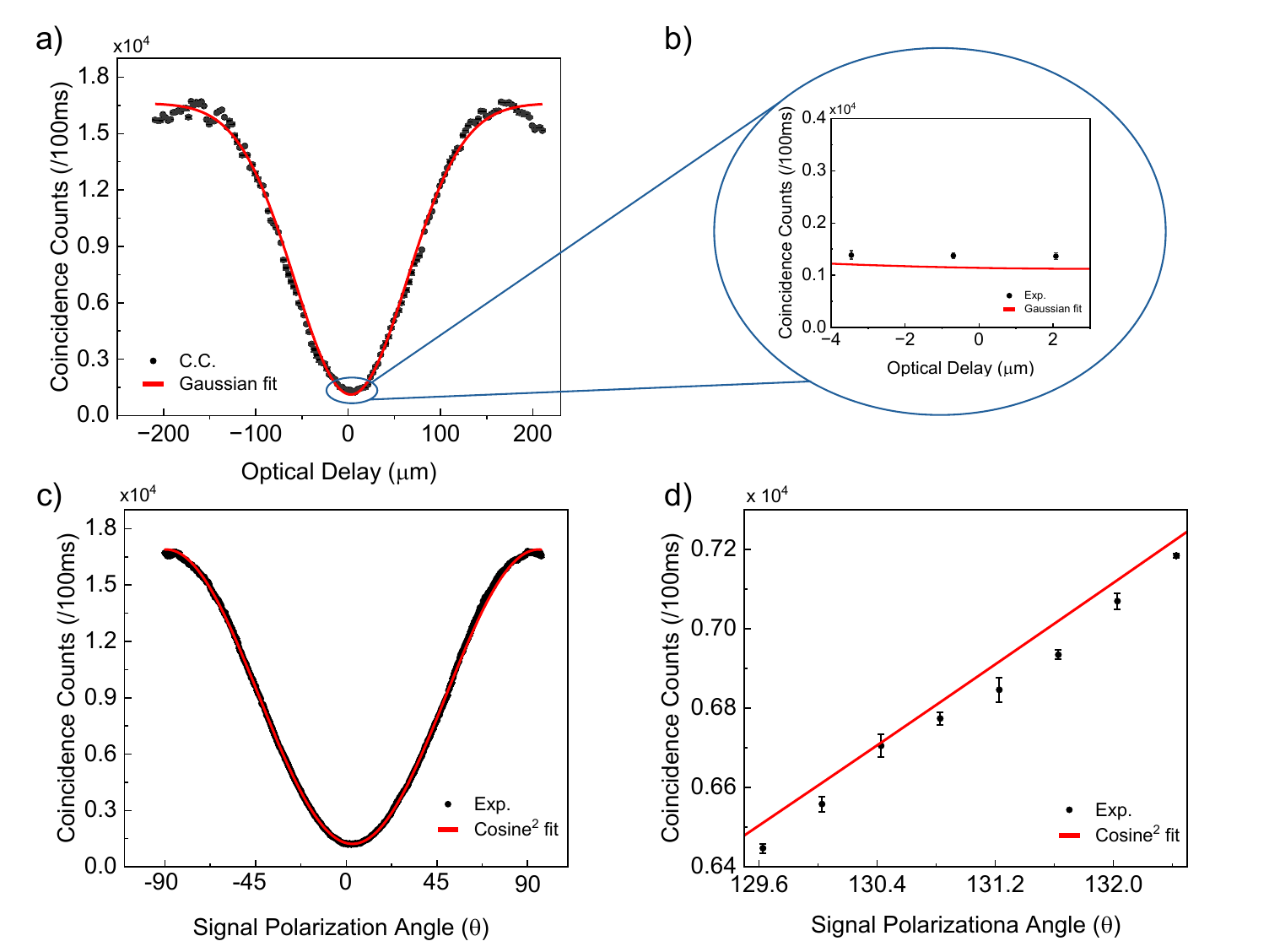}
    \caption{(a) Coincidence counts as a function of the relative optical delay between a pair of photons with identical polarization states, showing the characteristic HOM interference dip. (b) Magnified view of the HOM dip near zero relative optical delay. (c) Coincidence counts as a function of the signal-photon polarization angle for a fixed horizontal polarization of the idler photon, with the relative optical delay maintained at zero. (d) Enlarged view of the linear region of the polarization-dependent HOM curve, illustrating the polarization sensitivity and resolution of the measurement. The solid red lines represent fits to the experimental data. 
    }
    \label{Fig_4}
\end{figure}
We characterized the HOM interference in both the temporal-delay and polarization domains, with the results presented in Fig. \ref{Fig_4}. For the temporal HOM measurement, polarization indistinguishability between the signal and idler photons is ensured by aligning the fast axes of both the vortex waveplate (V) in the idler arm and the half-wave plate (P4) in the signal arm along the vertical direction, thereby leaving the polarization states of the photons unchanged. The relative optical delay between the photon pair is then varied by changing the optical path length of the signal photon while keeping that of the idler photon fixed. The coincidence counts at the two output ports of the beam splitter (BS), subsequently recorded as a function of the relative delay, are shown in Fig. \ref{Fig_4}(a). 
As evident from Fig. \ref{Fig_4}(a), the coincidence counts as a function of the relative optical delay exhibit the characteristic HOM interference profile, with a minimum coincidence count occurring at zero relative delay, where the signal and idler photons become temporally indistinguishable. 
The FWHM of the interference dip, obtained from a Gaussian fit (solid red line), is measured to be (144.6 $\pm$ 1.3 $\upmu$m) with an interference visibility of ($\sim$94$\%$). Such a relatively broad HOM dip is intentionally employed in the present experiment to minimize systematic errors in the polarization measurements arising from residual offsets and fluctuations in the optical delay (see Section \ref{sec_2.3}). The increased dip width results from the narrow spectral bandwidth of the detected photons, determined by the transmission spectrum of the interference filter with FWHM bandwidth of $\sim$ 3.2 nm \cite{singh2023near,singh2024fast}.
To provide a clearer perspective, the HOM interference curve near zero relative delay is magnified in Fig. \ref{Fig_4}(b). As evident, the coincidence counts exhibit negligible variation over an optical-delay range of approximately (6 $\upmu$m), which is sufficiently large to suppress the sensitivity of the interferometer to optical-path-length variations arising from transverse thickness variations of the sample as well as fluctuations in the relative delay between the two interferometer arms.
\\
On the other hand, to measure the HOM interference with the relative rotation of polarization state of the pair photon, we locked the delay stage to ensure relative optical delay between the pair photons to $x = 0$. Keeping the experimental setup unchanged, we rotated the fast axis angle of the HWP (P4) with respect to the vertical by the high-precision motorized rotation stage while maintaining the vertical polarization for idler photons. As evident from Fig. \ref{Fig_4}(c), the coincidence counts between the BS output ports show a characteristic HOM curve, with minimum coincidence for vertical polarization of the signal and indistinguishability with the idler polarization, the sign convention for the polarization angle follows Sec.~\ref{sec_2.2} . 
The experimental data (black dots) closely follow a $\mathrm{cosine}^{2}$ fit (solid red curve), confirming that the dependence of the coincidence counts on the signal polarization angle is consistent with the theoretical prediction of ($P_c$) given by Eq. \ref{eq_9}. To highlight the sensitivity of the HOM interferometer to relative polarization rotations, the linear region of the curve is magnified in Fig. \ref{Fig_4}(d). A linear fit (red line) to the experimental data yields a slope of approximately $\sim$262 counts/sec/degree, with clearly distinguishable data points corresponding to polarization-angle increments of (0.4$^\circ$). These results indicate that the interferometer can resolve polarization changes as small as ($\sim$0.4$^\circ$). The achievable resolution is primarily limited by the combined uncertainty arising from waveplate birefringence tolerances \cite{baur2021}, detector noise, and timing jitter \cite{hernandez2017}.

\subsection{Polarization projection using HOM interference}
\begin{figure}[h]
\centering
\includegraphics[width=0.9\linewidth]{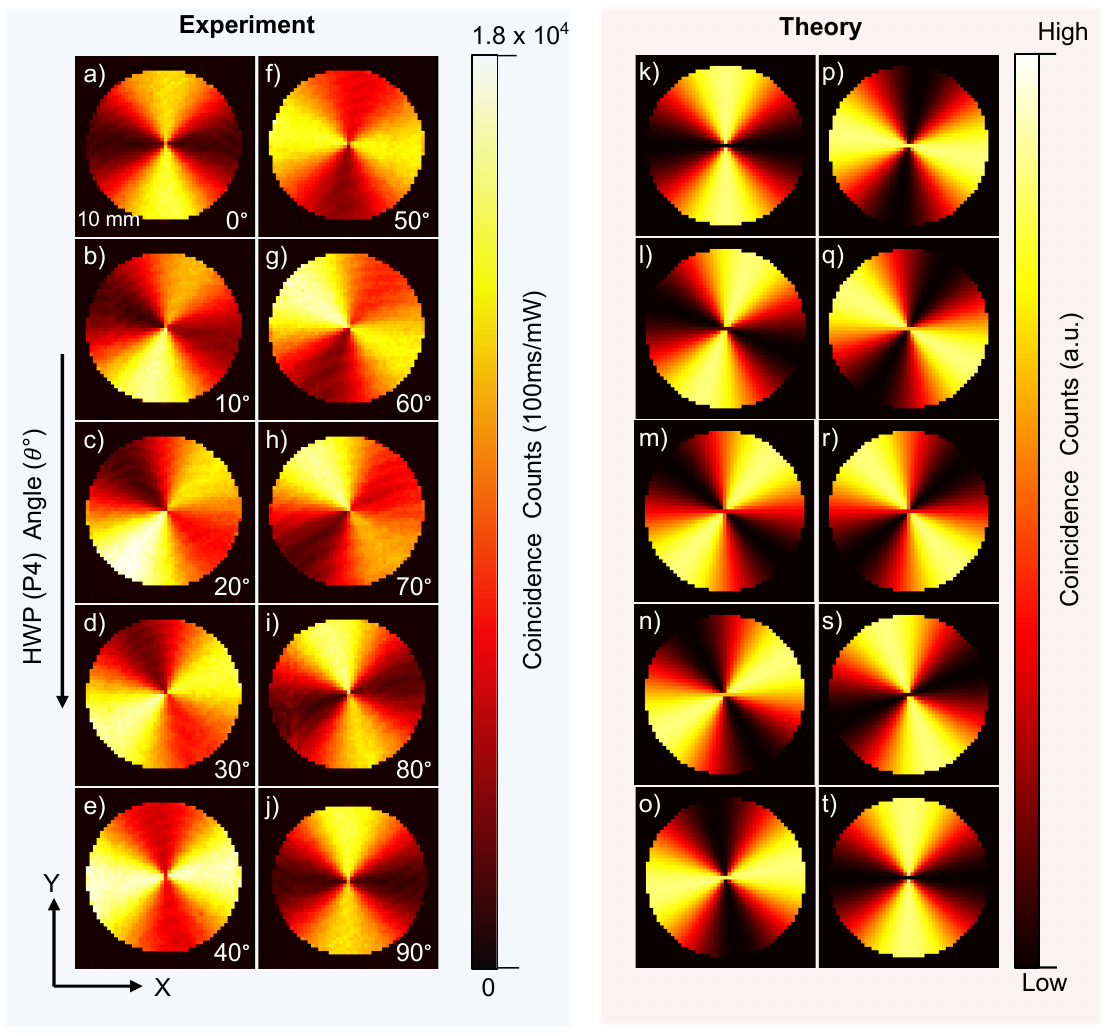}
\caption{(a–j) Experimental and (k–t) theoretical reconstructed HOM coincidence-count distributions representing the transverse polarization distribution of the vortex waveplate for different angles of HWP (P4) on the signal-photons.}
\label{Fig_5}
\end{figure}

Following the complete characterization of the HOM interferometer in terms of its sensitivity to polarization and optical delay, we investigated the transverse polarization distribution of idler photons propagating through a vortex waveplate of order (m = 1), possessing an azimuthally varying optic-axis orientation. Using the experimental setup shown in Fig. \ref{Fig_2}, we first adjusted the optical delay and polarization states of the pair photons such that the coincidence counts are minimum (see Fig. \ref{Fig_4}(c)). 
Keeping the remaining experimental parameters fixed, the vortex waveplate mounted on the computer-controlled two-dimensional translation stage, with 50 nm minimum step size, is scanned in a raster pattern across its transverse plane with a step size of 0.5 mm. The $1/e^{2}$ idler Gaussian beam diameter at the input aperture of vortex waveplate was $\sim$ 0.325 $\pm$ 0.013 mm, smaller than the 0.5 mm scan step, so the adjacent raster positions probe distinct, non-overlapping and localized regions of the optic-axis distribution. Consequently, the idler photons traversed different regions of the waveplate aperture, experiencing spatially varying fast-axis orientations and the associated polarization rotations. The resulting polarization rotation introduces distinguishability between the fixed-polarization (vertical) signal photons and the polarization-rotated idler photons, leading to variations in the HOM coincidence counts. Owing to the high brightness of the photon-pair source, coincidence counts are recorded with an integration time of 100 ms and averaged over 1 s for each scan position. By rotating the signal-photon polarization by adjusting the HWP (P4) angle in steps of (10$^\circ$), we experimentally measured and theoretically calculated the corresponding transverse coincidence-count distributions, with the results shown in Fig. \ref{Fig_5}. The Fig. \ref{Fig_5}(a–j) shows the experimentally reconstructed coincidence-count distributions arising from HOM interference across the transverse aperture of the vortex waveplate for different signal-photon polarization angles. For each signal-photon polarization angle varied through the rotation of HWP (P4) angle $\theta$ in steps of 10$^\circ$, the raster-scan procedure was repeated, and the corresponding transverse polarization map was reconstructed from the measured coincidence counts. Each pixel corresponds to a unit step in the $xy$ raster scan. As expected for a vortex waveplate of order, m = 1, the reconstructed patterns exhibit two lobes separated by low-coincidence regions arising from photon bunching, closely resembling the classical polarization-projection images shown in Fig. \ref{Fig_3}. It should be noted, however, that the classical measurements employed a larger beam size to illuminate a substantial portion of the vortex waveplate and reveal the full azimuthal variation of its fast-axis orientation, resulting in the characteristic doughnut-shaped intensity profile. In contrast, the present HOM-based measurements employed a smaller beam size to locally probe the fast-axis orientation and thereby avoid spatial averaging over regions with different optic-axis directions. 

Similarly, the theoretically reconstructed coincidence-count distributions for the corresponding signal-photon polarization angles, obtained using Eq.~\ref{eq_9}, are shown in Fig. \ref{Fig_5}(k–t). At each raster position $(x', y')$, the local azimuthal coordinate on the vortex waveplate is $\phi(x',y') = \tan^{-1}\!\left((y'-y_0)/(x'-x_0)\right)$, where $(x_0,y_0)$ denotes the location of the center of the plate vortex waveplate.  A vertically polarized idler photon traversing the vortex waveplate, acquires a polarization rotation that, relative to the signal polarization fixed by the HWP (P4) at angle $\theta_{\mathrm{P4}}$, yields a relative polarization angle given as
\begin{equation}
\theta(x',y') = 2\!\left(\theta_{\mathrm{f}}(x',y)' - \theta_{\mathrm{P4}}\right) = m\,\phi(x',y') + 2\delta - 2\theta_{\mathrm{P4}}.
\label{eq_15}
\end{equation}
Substituting Eq.~\ref{eq_15} into Eq.~\ref{eq_9} at zero relative optical delay ($x = 0$) gives the reconstructed coincidence-count distribution,
\begin{equation}
N_c(x',y';\theta_{\mathrm{P4}}) = N_{\max}\!\left(\,1 - \alpha\,\cos^{2}\!\big(m\,\phi(x',y') + 2\delta - 2\theta_{\mathrm{P4}}\big)\right),
\label{eq_16}
\end{equation}
where $N_{\max}$ is the mean count scale set by the source brightness and the integration time. Equation~\ref{eq_16} reproduces the two-lobed structure expected for $m=1$, and stepping $\theta_{\mathrm{P4}}$ in $10^\circ$ increments rotates the pattern by $(20^\circ)$ in the transverse plane, consistent with the experimental panels of Fig. ~\ref{Fig_5}. The close agreement between theory and experiment is evident from the similarity of the reconstructed patterns. Nevertheless, the low-coincidence regions are more pronounced in the theoretical images than in the experimental results. This discrepancy can be attributed primarily to the residual birefringence of the vortex waveplate, which was designed for operation at 1064 nm but is used here at 810 nm, leading to incomplete polarization transformation. In addition, the experimental images were reconstructed directly from the raw coincidence data without applying any background correction based on the minimum coincidence counts of the HOM interferometer (see Fig. \ref{Fig_4}). Incorporating these effects into the model is expected to further improve the agreement between theory and experiment. To quantify the correspondence between the reconstructed images, we evaluated the overlap fidelity between the theoretical and experimental patterns, obtaining an average fidelity of approximately (95$\%$) \cite{cai2025}. These results demonstrate that HOM interference can be employed to directly map spatially varying polarization distributions through coincidence measurements without requiring post-sample polarization projections.

\subsection{Extracting polarization orientation and birefringence pattern}

\begin{figure}[h]
    \centering
    \includegraphics[width=\linewidth]{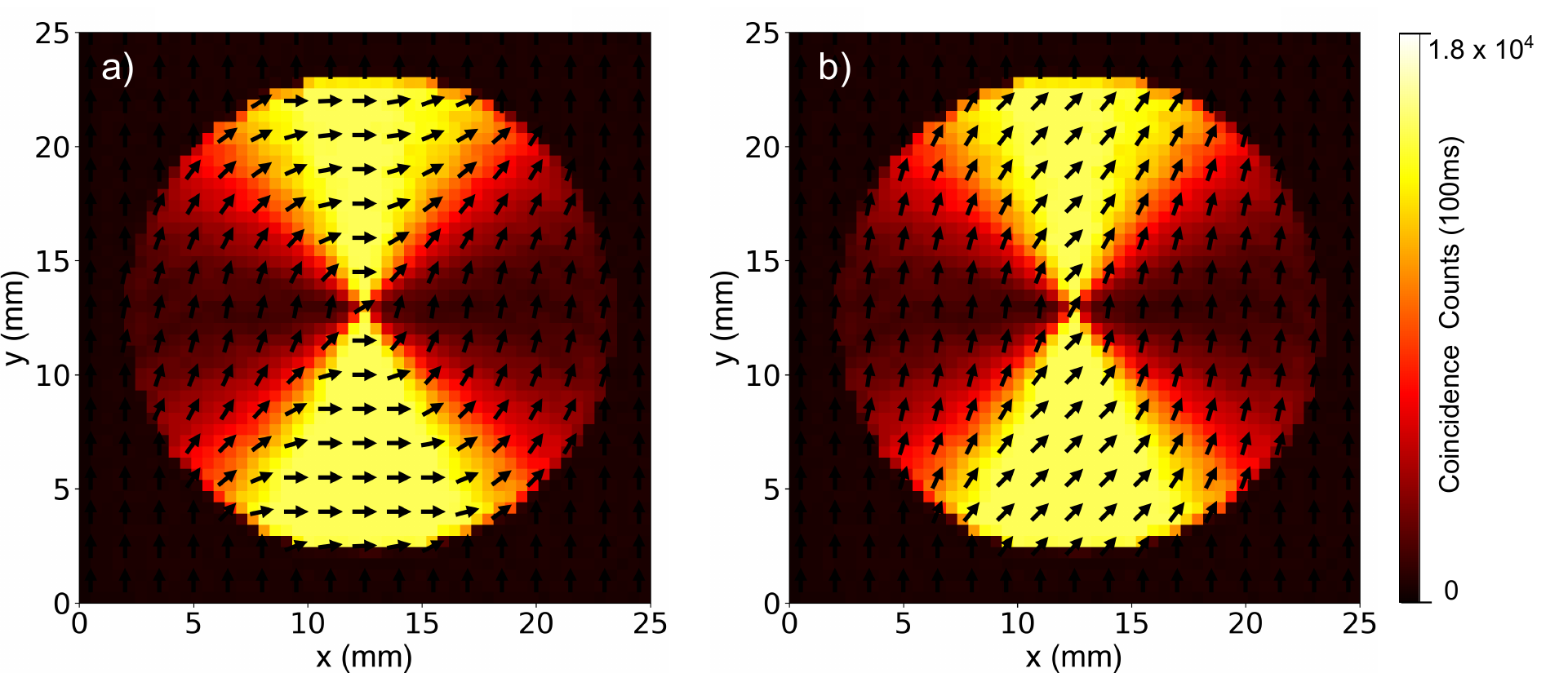}
    \caption{Reconstructed HOM coincidence-count distributions representing the transverse distributions of the (a) polarization-axis orientation of the idler photons and (b) the local fast-axis orientation of the vortex waveplate.The black arrows the direction.}
    \label{Fig_6}
\end{figure}

We further extracted the polarization orientation of the idler photons arising from the spatially varying birefringence characteristics of the vortex waveplate. While each image of Fig. \ref{Fig_5}(a-j) carries the signature of the polarization orientation, for illustration, we consider the image of Fig. \ref{Fig_5}(a). By using the established relationship between $P_{c}$ and $\theta$ from Eq. \ref{eq_9}, in conjunction with their calibration parameters derived from Fig. \ref{Fig_4}(c), we estimated the polarization orientation of the idler photons, denoted by the black arrows in Fig. \ref{Fig_6}(a). 
However, because HOM interference depends only on the magnitude of the polarization overlap between the signal and idler photons and is insensitive to the sign (direction) of the polarization vector (see $\theta$ in Eq. \ref{eq_9}) as well as any associated optical phase, the reconstructed black arrows indicate the local polarization orientation axis but not the actual orientation of the polarization. This leads to a fundamental two‑fold ambiguity between angles $\theta$ and ($180^{\circ} - \theta$), producing identical visibility for both cases. While it prevents the discrimination of these two angles and restricts the physically meaningful measurement range to $[0^{\circ}, 90^{\circ}]$, all other angles get folded into this interval. The arrows in Fig. \ref{Fig_6}(a) therefore represent the polarization axis with the acute angle relative to the vertical direction (signal is vertical for image Fig. \ref{Fig_5}(a)). The regions of high coincidence counts (top and bottom) indicate no bunching, implying that the idler photons are polarized orthogonally to the vertically polarized signal photons. This means the idler polarization is orientation $\theta \sim 90^{\circ}$, consistent with the arrow directions. 
However, these regions correspond to the extrema of the $\cos^{2}\theta$ response, where the sensitivity vanishes since $dP_c/d\theta=0$ (see Eq.~\ref{eq_9}). The same argument applies to the low-coincidence region around $\theta \approx 0^{\circ}$. Consequently, small angular deviations modify the coincidence probability only through the second-order term in $\delta\theta$, making them intrinsically difficult to resolve. Furthermore, the two extrema differ in their shot-noise characteristics. Near the high-coincidence maximum ($\theta \approx 90^{\circ}$), the large mean coincidence count $N$ is accompanied by a Poisson fluctuation of $\sqrt{N}$, against which the second-order signal must be detected. In contrast, near the low-coincidence minimum ($\theta \approx 0^{\circ}$), the small value of $N$ results in a large relative shot-noise uncertainty, $\sqrt{N}/N = 1/\sqrt{N}$, which likewise obscures the second-order signal. Here, $N$ denotes the mean coincidence count within the measurement window.
\par

\noindent On the other hand, the minimum coincidence counts regions (left and right), which are expected to exhibit nearly zero coincidence counts, do not correspond to ($\theta \approx 0^\circ$), indicating that the idler polarization axis at those spatial locations deviates from the expected vertical orientation. This observation is consistent with the non-vertical polarization axes indicated by the arrows in Fig. \ref{Fig_6}(a). The deviation arises from residual coincidence counts at the minimum of the HOM dip, corresponding to a finite angular mismatch ($\Delta\theta$) between the signal and idler polarizations. From the measured coincidence counts, we estimate ($\Delta\theta \approx 9^\circ$) in the low-count regions, in good agreement with the reconstructed polarization orientations. This offset can be attributed to several experimental imperfections. In particular, the vortex waveplate was designed for operation at 1064 nm but used at 810 nm, resulting in wavelength-dependent retardance errors that can introduce ellipticity into the idler polarization and thereby reduce its overlap with the vertically polarized signal photon. Furthermore, spatial non-uniformity of the retardance and local deviations of the fast-axis orientation from the ideal design of the vortex waveplate can introduce additional polarization errors. Collectively, these effects prevent the idler polarization from attaining the ideal vertical polarization state in regions where the coincidence counts are expected to be minimum, leading to the observed tilt of the polarization axes in the reconstructed map.

\noindent Using the results of Fig. \ref{Fig_6}(a) we have estimated the spatial distribution of the fast axis orientation of the vortex waveplate using the relation $\phi = \theta/2$. Here, $\theta$ is the polarization angle of the idler photons inferred from Fig. \ref{Fig_6}(a). This factor of one‑half arises because a half-wave retarder rotates the polarization by twice the angle between the input polarization and its fast axis \cite{saleh2019}. 
The Fig. \ref{Fig_6}(b) displays the resulting birefringence, thus the fast-axis orientation map of the vortex waveplate, with arrows in black indicating the local fast‑axis direction. On similar lines, regions corresponding to high coincidence counts (top and bottom) yield birefringence angles near $45^\circ$ (as $\theta \sim 90^\circ$), whereas the low‑count regions (left and right) show angles clustering around $\sim4.5^\circ$. 
The birefringence map thus provides a complementary visualization of the spatially varying optical properties of the vortex waveplate and confirms that the device does not achieve perfect half‑wave retardation everywhere at the operating wavelength.\

\noindent It is worth noting that, in addition to imperfections of the vortex waveplate, the overall measurement precision is influenced by several systematic effects. In particular, spatially resolved polarization measurements can be affected by optical-delay-induced polarization errors, as discussed in Sec. \ref{sec_2.2}. For a representative optical-delay variation of ($\lambda$/4) ($\lambda \approx$ 810 nm) and an HOM dip width of ($\sigma_x \approx$ 144 $\upmu$m), the resulting spurious polarization error is estimated to be ($\delta\theta \approx 0.08^\circ$), which sets the fundamental precision limit under ideal operating conditions. In principle, this limit can be further reduced by increasing the HOM dip width, owing to the inverse scaling ($\delta\theta \propto 1/\sigma_x$). This may be achieved through narrower spectral filtering or by employing a type-II SPDC source with longer coherence length. However, increasing the HOM width generally reduces the available photon flux and coincidence-to-accidental ratio (CAR), thereby increasing the acquisition time required to achieve a given signal-to-noise ratio. Compensating for the reduced pair rate by increasing the pump power is also undesirable, as it enhances multi-photon contributions that degrade HOM visibility. Consequently, improving the polarization precision through broader HOM interference dips is ultimately constrained by a trade-off between precision, brightness, acquisition time, and interference visibility.

\section{Conclusion}

In conclusion, we have demonstrated a projection-free method for mapping transverse polarization distributions using polarization-dependent Hong–Ou –Mandel (HOM) interference. Leveraging the sensitivity of two-photon interference to polarization distinguishability, we have directly mapped the spatially varying polarization transformations introduced by a vortex waveplate onto coincidence counts without the need for post-sample polarization projections. Using a high-brightness SPDC photon-pair source, we achieved rapid data acquisition and reconstructed the transverse polarization distribution with an average fidelity of approximately 95$\%$ and an angular resolution of about (0.4$^\circ$). The reconstructed polarization maps showed excellent agreement with theoretical predictions, with small discrepancies attributed primarily to wavelength-dependent retardance errors and non-idealities of the vortex waveplate. We further analyzed the fundamental precision limits arising from residual optical-delay fluctuations and discussed the trade-offs between measurement precision, HOM-dip width, photon flux, and interference visibility. The proposed technique eliminates the need for sequential polarization projections and associated calibration errors, providing a simple, scalable, and high-precision quantum-optical approach for characterizing structured polarization fields in birefringent materials. The current results highlight the broader potential of HOM interferometry as a versatile platform for quantum sensing and imaging applications without the need for polarization projection systems.



\subsection{Acknowledgments}
The authors acknowledge the support of the Department of Space, Govt. of India.

\subsection{Conflicts of Interest}

The authors declare no conflicts of interest.

\subsection{Data Availability Statement}
The data supporting the findings of this study are available from the corresponding author upon reasonable request.

\subsection{Supporting Information}
No supplementary documents are available for this study.
\bibliography{mybibliography}

@article{mndlovu2024review,
  title={A review of bio material degradation assessment approaches employed in the biomedical field},
  author={Mndlovu, Hillary and Kumar, Pradeep and du Toit, Lisa C and Choonara, Yahya E},
  journal={npj Materials Degradation},
  volume={8},
  number={1},
  pages={66},
  year={2024},
  nolink = {},
  publisher={Nature Publishing Group UK London}
}

@article{Jabir:2017,
  title={Robust, high brightness, degenerate entangled photon source at room temperature},
  author={Jabir, M. V. and Samanta, G. K.},
  journal={Sci. Reports},
  volume={7},
  number={1},
  pages={12613},
  nolink = {},
  year={2017},
  }

@article{he2021polarisation,
  title={Polarisation optics for biomedical and clinical applications: a review},
  author={He, Chao and He, Honghui and Chang, Jintao and Chen, Binguo and Ma, Hui and Booth, Martin J},
  journal={Light: Science \& Applications},
  volume={10},
  number={1},
  pages={194},
  year={2021},
  nolink = {},
  publisher={Nature Publishing Group UK London}
}

@article{tan2024,
  title={Real-time polarization compensation method in quantum communication based on channel Muller parameters detection},
  author={Tan, Yongjian and Wang, Jianyu and Wu, Jincai and He, Zhiping},
  journal={Communications Engineering},
  volume={3},
  number={1},
  pages={57},
  year={2024},
  nolink = {},
  publisher={Nature Publishing Group UK London}
}

@inproceedings{baur2021,
  title={Performance limitations in optical retarders},
  author={Baur, Tom and Kraemer, Michael},
  booktitle={Polarization Science and Remote Sensing X},
  volume={11833},
  pages={111--124},
  year={2021},
  nolink = {},
  organization={SPIE}
}

@article{singh2023near,
  title={Near-Video Frame Rate Quantum Sensing Using Hong--Ou--Mandel Interferometry},
  author={Singh, Sandeep and Kumar, Vimlesh and Sharma, Varun and Faccio, Daniele and Samanta, GK},
  journal={Advanced Quantum Technologies},
  volume={6},
  number={11},
  pages={2300177},
  year={2023},
  nolink = {},
  publisher={Wiley Online Library}
}

@article{Hong1987Mandel,
  title={Measurement of subpicosecond time intervals between two photons by interference},
  author={Hong, C. K. and Ou, Z. Y. and Mandel, Leonard},
  journal={Physical Review Letters},
  volume={59},
  number={18},
  pages={2044--2046},
  year={1987},
  nolink = {},
  publisher={American Physical Society}
}

@article{singh2023tolerance,
  title={A Tolerance-Enhanced Spontaneous Parametric Downconversion Source of Bright Entangled Photons},
  author={Singh, Sandeep and Kumar, Vimlesh and Ghosh, Anirban and Forbes, Andrew and Samanta, Goutam K},
  journal={Advanced Quantum Technologies},
  volume={6},
  number={2},
  pages={2200121},
  year={2023},
  nolink = {},
  publisher={Wiley Online Library}
}

@article{singh2024fast,
  title={Fast measurement of group index variation with optimum precision using Hong--Ou--Mandel interferometry},
  author={Singh, Sandeep and Kumar, Vimlesh and Samanta, GK},
  journal={APL Quantum},
  volume={1},
  number={4},
  year={2024},
  nolink = {},
  publisher={AIP Publishing}
}

@article{polarization2020,
  title={Tracking the polarisation state of light via Hong-Ou-Mandel interferometry},
  author={Harnchaiwat, Natapon and Zhu, Feng and Westerberg, Niclas and Gauger, Erik and Leach, Jonathan},
  journal={Optics express},
  volume={28},
  number={2},
  pages={2210--2220},
  year={2020},
  nolink = {},
  publisher={Optica Publishing Group}
}

@article{hernandez2017,
  title={A computational model of a single-photon avalanche diode sensor for transient imaging},
  author={Hernandez, Quercus and Gutierrez, Diego and Jarabo, Adrian},
  journal={arXiv preprint arXiv:1703.02635},
  nolink = {},
  year={2017}
}

@article{tominaga2008,
  title={Polarization imaging for material classification},
  author={Tominaga, Shoji and Kimachi, Akira},
  journal={Optical Engineering},
  volume={47},
  number={12},
  pages={123201--123201},
  year={2008},
  nolink = {},
  publisher={Society of Photo-Optical Instrumentation Engineers}
}

@article{ghosh2011,
  title={Tissue polarimetry: concepts, challenges, applications, and outlook},
  author={Ghosh, Nirmalya and Vitkin, I Alex},
  journal={Journal of biomedical optics},
  volume={16},
  number={11},
  pages={110801--110801},
  year={2011},
  nolink = {},
  publisher={Society of Photo-Optical Instrumentation Engineers}
}

@article{kumar2023,
  title={Controlling the coverage of full Poincar{\'e} beams through second-harmonic generation},
  author={Kumar, Subith and Saripalli, Ravi K and Ghosh, Anirban and Buono, Wagner T and Forbes, Andrew and Samanta, GK},
  journal={Physical Review Applied},
  volume={19},
  number={3},
  pages={034082},
  year={2023},
  nolink = {},
  publisher={APS}
}

@article{han2023,
  title={Attosecond metrology in circular polarization},
  author={Han, Meng and Ji, Jia-Bao and Ueda, Kiyoshi and W{\"o}rner, Hans Jakob},
  journal={Optica},
  volume={10},
  number={8},
  pages={1044--1052},
  year={2023},
  nolink = {},
  publisher={Optica Publishing Group}
}

@article{zhang2024,
  title={Quantum imaging of biological organisms through spatial and polarization entanglement},
  author={Zhang, Yide and He, Zhe and Tong, Xin and Garrett, David C and Cao, Rui and Wang, Lihong V},
  journal={Science Advances},
  volume={10},
  number={10},
  pages={eadk1495},
  nolink = {},
  year={2024},
  nolink = {},
  publisher={American Association for the Advancement of Science}
}

@article{waldchen2015,
  title={Light-induced cell damage in live-cell super-resolution microscopy},
  author={W{\"a}ldchen, Sina and Lehmann, Julian and Klein, Teresa and Van De Linde, Sebastian and Sauer, Markus},
  journal={Scientific reports},
  volume={5},
  number={1},
  pages={15348},
  year={2015},
  nolink = {},
  publisher={Nature Publishing Group UK London}
}

@article{dyer2009,
  title={High-brightness, low-noise, all-fiber photon pair source},
  author={Dyer, Shellee D and Baek, Burm and Nam, Sae Woo},
  journal={Optics express},
  volume={17},
  number={12},
  pages={10290--10297},
  year={2009},
  nolink = {},
  publisher={Optical Society of America}
}

@article{gonccalves2025,
  title={Quantum Imaging of Birefringent Samples using Hong-Ou-Mandel Interference},
  author={Gon{\c{c}}alves, Carolina and Ferreira, Tiago D and Monteiro, Catarina S and Silva, Nuno A},
  journal={arXiv preprint arXiv:2512.19637},
  nolink = {},
  year={2025}
}

@book{saleh2019,
  title={Fundamentals of photonics, 2 volume set},
  author={Saleh, Bahaa EA and Teich, Malvin Carl},
  year={2019},
  nolink = {},
  publisher={john Wiley \& sons}
}

@article{cai2025,
  title={Rapid diffused optical imaging for accurate 3D estimation of subcutaneous tissue features},
  author={Cai, Shanshan and Mai, John and Hong, Winn and Fraser, Scott E and Cutrale, Francesco},
  journal={Iscience},
  volume={28},
  number={2},
  year={2025},
  nolink = {},
  publisher={Elsevier}
}

@book{goldstein2017,
  title={Polarized light},
  author={Goldstein, Dennis H},
  year={2017},
  nolink = {},
  publisher={CRC press}
}

@inproceedings{singh2021,
  title={Variation of the Hong-Ou-Mandel interference dip with crystal length},
  author={Singh, Sandeep and Sharma, Varun and Kumar, Vimlesh and Samanta, GK},
  booktitle={European Quantum Electronics Conference},
  pages={eb\_p\_26},
  year={2021},
  nolink = {},
  organization={Optica Publishing Group}
}

@inproceedings{singh2023,
  title={Hong-Ou-Mandel interferometry for high precision sensing of real-time vibrations},
  author={Singh, Sandeep and Kumar, Vimlesh and Sharma, Varun and Faccio, Daniele and Samanta, GK},
  booktitle={Frontiers in Optics},
  pages={JTu5A--58},
  year={2023},
  nolink = {},
  organization={Optica Publishing Group}
}

@article{lyons2018,
  title={Attosecond-resolution hong-ou-mandel interferometry},
  author={Lyons, Ashley and Knee, George C and Bolduc, Eliot and Roger, Thomas and Leach, Jonathan and Gauger, Erik M and Faccio, Daniele},
  journal={Science advances},
  volume={4},
  number={5},
  pages={eaap9416},
  nolink = {},
  year={2018},
  
  publisher={American Association for the Advancement of Science}
}

@article{cardano2012,
  title={Polarization pattern of vector vortex beams generated by q-plates with different topological charges},
  author={Cardano, Filippo and Karimi, Ebrahim and Slussarenko, Sergei and Marrucci, Lorenzo and De Lisio, Corrado and Santamato, Enrico},
  journal={Applied optics},
  volume={51},
  number={10},
  pages={C1--C6},
  year={2012},
  nolink = {},
  publisher={Optical Society of America}
}

@article{kadiri2019,
  title={Wavelength-adaptable effective q-plates with passively tunable retardance},
  author={Kadiri, Gururaj and Raghavan, G and others},
  journal={Scientific reports},
  volume={9},
  number={1},
  pages={11911},
  year={2019},
  nolink = {},
  publisher={Nature Publishing Group}
}

@article{caspani2017,
  title={Integrated sources of photon quantum states based on nonlinear optics},
  author={Caspani, Lucia and Xiong, Chunle and Eggleton, Benjamin J and Bajoni, Daniele and Liscidini, Marco and Galli, Matteo and Morandotti, Roberto and Moss, David J},
  journal={Light: Science \& Applications},
  volume={6},
  number={11},
  pages={e17100--e17100},
  year={2017},
  nolink = {},
  publisher={Springer Science and Business Media LLC}
}

@misc{mandel1996,
  title={Optical coherence and quantum optics},
  author={Mandel, Leonard and Wolf, Emil and Shapiro, Jeffrey H},
  year={1996},
  nolink = {},
  publisher={American Institute of Physics}
}

@article{karimi2007,
  title={Hypergeometric-gaussian modes},
  author={Karimi, Ebrahim and Zito, Gianluigi and Piccirillo, Bruno and Marrucci, Lorenzo and Santamato, Enrico},
  journal={Optics letters},
  volume={32},
  number={21},
  pages={3053--3055},
  nolink = {},
  year={2007},
  publisher={Optical Society of America}
}

@article{nabadda2024,
  title={Mueller matrix imaging polarimeter with polarization camera self-calibration applied to structured light components},
  author={Nabadda, Esther and S{\'a}nchez-L{\'o}pez, Mar{\'\i}a del Mar and Vargas, Asticio and Lizana, Angel and Campos, Juan and Moreno, Ignacio},
  journal={Journal of the European Optical Society-Rapid Publications},
  volume={20},
  number={1},
  pages={5},
  year={2024},
  nolink = {},
  
  publisher={EDP Sciences}
}

@article{zhi2017,
  title={Error analysis and Stokes parameter measurement of rotating quarter-wave plate polarimeter},
  author={Zhi, Dan-Dan and Li, Jian-Jun and Gao, Dong-Yang and Zhai, Wen-Chao and Huang, Xiong-Hao and Zheng, Xiao-Bing},
  journal={Chinese Physics B},
  volume={26},
  number={12},
  pages={124201},
  year={2017},
  nolink = {},
  publisher={Chinese Physical Society and IOP Publishing Ltd}
}

@article{kumar2022,
  title={Imaging inspired characterization of single photons carrying orbital angular momentum},
  author={Kumar, Vimlesh and Sharma, Varun and Singh, Sandeep and Kumar, S Chaitanya and Forbes, Andrew and Ebrahim-Zadeh, Majid and Samanta, Goutam K},
  journal={AVS Quantum Science},
  volume={4},
  number={1},
  year={2022},
  nolink = {},
  publisher={AIP Publishing}
}

@inproceedings{singh2022generation,
  title={Generation of down-converted photons in a “perfect” annular ring insensitive to the change in phase-matching},
  author={Singh, Sandeep and Kumar, Vimlesh and Ghosh, Anirban and Samanta, GK},
  booktitle={Laser Science},
  pages={JW5A--29},
  year={2022},
  nolink = {},
  organization={Optica Publishing Group}
}
\bibliographystyle{quantum}

\end{document}